\documentclass[final,1p,times,twocolumn]{elsarticle}
\usepackage{epsfig}
\usepackage{amssymb}
\usepackage{amsmath}
\usepackage{subcaption}
\usepackage{longtable}
\journal{Nuclear Physics B}
\usepackage{xcolor}

\DeclareMathAlphabet{\oldcal}{OMS}{cmsy}{m}{n}
\newcommand{\bigo}[1]{\oldcal{O}\left(#1\right)}
\usepackage{hyperref}
\usepackage{multirow}
\usepackage{float}

\usepackage[colorinlistoftodos]{todonotes}

\begin{document}

\title{Six-loop $\varepsilon$ expansion of three-dimensional $\text{U}(n)\times \text{U}(m)$ models}
\author[label1]{L.\,Ts.\,Adzhemyan}
\author[label2]{E.~V. Ivanova}
\author[label1]{M.\,V.\,Kompaniets}
\author[label3]{\corref{cor1}A.\,Kudlis}
\ead{andrewkudlis@gmail.com} 
\cortext[cor1]{Corresponding author}
\author[label1]{{\textdagger}A.\,I.\,Sokolov}

\address[label1]{Saint Petersburg State University, 7/9 Universitetskaya Embankment, St. Petersburg, 199034 Russia}
\address[label2]{New Jersey Institute of Technology, 323 Dr Martin Luther King Jr Blvd, Newark, NJ 07102, USA}
\address[label3]{ITMO University, Kronverkskiy prospekt 49, Saint Petersburg 197101, Russia}

\date{\today}

\begin{abstract}
We analyze the Landau-Wilson field theory with $\text{U}(n)\times\text{U}(m)$ symmetry which describes the finite-temperature phase transition in QCD in the limit of vanishing quark masses with $n=m=N_f$ flavors and unbroken anomaly at the critical temperature. The six-loop expansions of the renormalization group functions are calculated within the Minimal Subtraction scheme in $4 - \varepsilon$ dimensions. The $\varepsilon$ series for the upper marginal dimensionality $n^{+}(m,4-\varepsilon)$ -- the key quantity of the theory -- are obtained and resummed by means of different approaches. The numbers found are compared with their counterparts obtained earlier within lower perturbative order. In particular, using an increase in the accuracy of numerical results for $n^{+}(m,3)$ by one order of magnitude, we strengthen the conclusions obtained within previous order in perturbation theory about fairness of the inequality $n^{+}(m,3)>m$. This, in turn, indicates the absence of a stable three-dimensional fixed point for $n=m$, and as a consequence a first-order kind of finite-temperature phase transition in light QCD.

\end{abstract}


\maketitle

\section{Introduction}

Understanding of the true nature of phase transitions in QCD is an important fundamental problem. It is currently believed that for hadronic matter at a low temperature, where the chiral symmetry is broken, there is a phase transition into a quark-gluon plasma associated with the restoration of chiral symmetry. How this transition occurs is essentially dependent on the QCD parameters, including quark masses and the number of flavors $N_f$. In particular, when the number of massless quarks is two or more, this transition is essentially related to chiral symmetry restoring~(for review, see, e.g., Ref.~\cite{Wilczek_arx_rev}).

The behavior of QCD with $N_f$ flavors of massless quarks is described by an action, which is classically invariant under the global flavor symmetry $\text{U}_A(1)\times \text{SU}(N_f)\times \text{SU}(N_f)$. The axial $\text{U}_A(1)$ symmetry can be quantum-mechanically violated, which is related to nonconservation of the corresponding current~\cite{pisarski1984}. In this case, the symmetry is reduced to $\text{Z}_A(N_f)\times \text{SU}(N_f)\times \text{SU}(N_f)$~\cite{pisarski1984}. At zero temperature, the symmetry is further spontaneously broken to $\text{SU}(N_f)$, while increasing the temperature leads to a phase transition associated with the restoration of the full chiral symmetry. In order to describe this phase transition, we consider the most general renormalisable $\text{U}(N_f)\times \text{U}(N_f)$ invariant action with $N_f$-by-$N_f$ complex matrices $\tilde{\varphi}$ as order parameters~\cite{pisarski1984,Wilczek_1992}:
\begin{equation}
S = \int d^D{x}\Bigg\{\frac{1}{2} \Bigg[\text{Tr}\,\partial_{\mu}\tilde{\varphi}^{\dagger}\partial_{\mu}\tilde{\varphi}^{}+ m_0^2\,\text{Tr}\,\tilde{\varphi}^{\dagger}\tilde{\varphi}^{}\Bigg] + \frac{1}{4!}\Bigg[ g_{1}^{0}\Big(\text{Tr}\,\varphi^{\dagger}\varphi^{}\Big)^2+g_{2}^{0}\text{Tr}\,\Big(\varphi^{\dagger}\varphi^{}\Big)^2 \Bigg]\Bigg\}, \label{H_int}
\end{equation}
This action provides the correct symmetry breaking pattern only for a positive bare coupling $g_{2}^{0}$. The range of stability of the action~\eqref{H_int} is determined by the conditions $g_{2}^{0}>0$ and $g_{1}^{0}+g_{2}^{0}/N_f>0$. If couplings satisfy them, then mean-field theory predicts a second-order phase transition. However, within the renormalization group (RG) approach, it is generally accepted that a continuous phase transition requires the existence of a stable fixed point (FP)~\cite{zinn1996book}, while its absence is considered as an indicator that the phase transition is first order~\cite{pelissetto2002critical}.

The RG analysis of the action~\eqref{H_int} was first performed about four decades ago~\cite{pisarski1984,pisarski1981,paterson1981} in $4-\varepsilon$ dimensions within the one-loop approximation. These works intended to find a stable FP in light QCD, which, however, was not found for all $N_f\geq2$. From a quantitative perspective, as experience shows, the conclusion about the type of phase transition often changes with an increase in order of perturbation theory. This kind of situation is observed, for example, in the cubic model~\cite{nelson1974renormalization, kleinert1995exact, shalaev1997five, kleinert1997stability, carmona2000n, adzhemyan2019six}. Thus, in order to determine the true nature of the phase transition, higher orders are required. Having realized this, the authors of Ref.~\cite{calabrese2004} performed a five-loop $\varepsilon$ analysis of the model~\eqref{H_int}. To be precise, they consider the more general $\text{U}(n)\times\text{U}(m)$ symmetric model, which allows one to get closer to the true structure of FPs for the original $\text{U}(N_f)\times\text{U}(N_f)$ model.
Since the detailed reasoning based on the one-loop approximation is given in~\cite{calabrese2004}, we only highlight some points. In general, when $n \neq m$, four FPs can appear. In contrast to the Gaussian FP $(0,0)$ and the Heisenberg FP $(g^*_H,0)$, which exist in the whole region of $(n,m)$, there are two more FPs, called $U^\pm$~\cite{calabrese2004}, that emerge in a bounded region of $n$ and $m$ values. The $U^{+}$ FP is a stable one and defines a new universality class. The corresponding area of stability is given by the inequality $n\geq n^+(m,d)$. Thus, if during the calculation, it turns out that the inequality $m \geq n^+(m,d)$ holds, then for the case $n=m=N_f$ one has a new universality class. Otherwise, it does not exist. It turned out that within the five-loop analysis, all calculations favor the inequality $m < n^+(m,d)$, and a first-order phase transition should be observed in light QCD. Apart from the $\varepsilon$ expansion, the six-loop approximation was obtained directly in $D=3$~\cite{butti2003}. This study saw no stable FP, which supports the conclusion about a first-order phase transition in model~\eqref{H_int}. 

This work's primary goal is to perform the RG calculations within the $\varepsilon$ expansion in the same high -- six-loop -- approximation, as was done in the RG in dimension $D=3$. These computations allow us to obtain highly accurate numerical estimates for the marginal dimensionality $n^{+}(m,3)$ that eliminates any doubts about the actual type of the phase transition predictable by the $\varepsilon$-expansion approach. This work is an extension of the six-loop RG expansions in terms of $\varepsilon$ for the basic model with $\text{O}(n)$ symmetry, which were obtained in Ref.~\cite{KP17}.

The paper is organized as follows. In Section~\ref{sec:model_renormalization} the $\text{U}(n)\times\text{U}(m)$ symmetric model and its renormalization are described. In Section~\ref{sec:rg_expansions} all the quantities of interest such as RG functions and marginal dimensionalities are defined. The numerical estimates of $n^{+}(m,3)$ for $m=\{2..6\}$ are presented in Section~\ref{sec:numerical_results}. Also, the numbers obtained in this work, found earlier within the lower-order approximation are compared. In the last section, a summary of the main results is presented.

\section{Model and renormalization}
\label{sec:model_renormalization}
\subsection{Model}
It is convenient from a technical point of view to rewrite the $\text{U}(n)\times\text{U}(m)$ symmetric model~\eqref{H_int} in the following form:
\begin{equation}
S = \int d^D{x}\Bigg\{\frac{1}{2} \Big[\partial_{\mu}\tilde{\varphi}_{i \alpha}^{\dagger}\partial_{\mu}\tilde{\varphi}_{\alpha i}^{}+ m_0^2\tilde{\varphi}_{i \alpha}^{\dagger}\tilde{\varphi}_{\alpha i}^{}\Big] + \frac{1}{4!}\Big[ g_{1}^{0}T^{(1)}_{\alpha i, \beta j, \gamma k, \delta l} +g_{2}^{0}T^{(2)}_{\alpha i,\beta j,\gamma k,\delta l} \Big] \tilde{\varphi}_{i\alpha}^{\dagger}\tilde{\varphi}_{j\beta}^{\dagger} \tilde{\varphi}_{\gamma k }^{}\tilde{\varphi}_{\delta l}^{} \Bigg\}, \label{H}
\end{equation}
where the bare field $\tilde{\varphi}$ is a complex matrix of size $n\times m$ $(\alpha\in\{1,\dots n\}$, $i\in\{1,\dots m\})$, $g_{1}^{0}$ and $g_{2}^{0}$ are bare coupling constants, $m_0$ is a bare mass being proportional to $T-T_0$, where $T_0$ is mean-field transition temperature. The tensor factors $T^{(1)}$ and $T^{(2)}$ are defined as follows:
\begin{eqnarray}
T^{(1)}_{\alpha i,\beta j,\gamma k, \delta l} &=& \frac{1}{2}\big[\delta_{\alpha i, \gamma k} \delta_{\beta j, \delta l} +\delta_{\alpha i, \delta l} \delta_{\beta j, \gamma l}\big], \quad \delta_{\alpha i, \beta j} = \delta_{\alpha \beta} \delta_{ij},\\
T^{(2)}_{\alpha i,\beta j,\gamma k, \delta l} &=& \frac{1}{2}\big[\delta_{\alpha i, \gamma l} \delta_{\beta j, \delta k} +\delta_{\beta j, \gamma l} \delta_{\alpha i, \delta k}\big].
\label{T12def}
\end{eqnarray}

As mentioned in the introduction, the difference of parameters $n$ and $m$ from each other leads to a very intricate RG flow structure. The $\text{U}(n)\times\text{U}(m)$ symmetric model with two coupling constants may possess four FPs. The Gaussian $(0,0)$ and Heisenberg $(g_H^*, 0)$ ones exist for all couples $(n,m)$. However, two others appear only in a limited range of parameter $n$ values. Similar to the chiral model ($\text{O}(n)\times\text{O}(m)$ symmetric one), there exist so-called marginal dimensionalities which separate the different regimes of critical behavior. Let us describe them. Following the notation suggested in~\cite{calabrese2004} one has:
\begin{enumerate}
    \item if $n<n^H(m,d)$ there are four FPs with $\text{O}(2mn)$ symmetric one being stable;
    \item if $n^H(m,d)<n<n^-(m,d)$ there are four FPs as well, but instead of $\text{O}(mn)$ symmetric 
    FP the $U^{+}$ one acquires the stability for negative $g_{2}^*$, for positive values of bare coupling $g^0_2$ the phase transition is expected to be first-order;
    \item for $n^-(m,d)<n<n^+(m,d)$ only Gaussian and Heisenberg FPs exist and, since both are unstable, the system is expected to undergo first-order phase transition for any values of bare couplings;
    \item if $n>n^+(m,d)$ there are again four FPs with $U^{+}$ one being stable, for which the coordinate $g_2^*$ is positive.
\end{enumerate}

\subsection{Renormalization}

The model ~\eqref{H} is multiplicatively renormalizable. The renormalization of parameters $g_{1}$, $g_{2}$, $m^2$ and field $\tilde{\varphi}$ is determined by the relations 
\begin{gather}
m_0^2= m^2 Z_{m^2}, \qquad  g^{0}_{1} = g^{}_1 \mu^{\varepsilon}Z_{g_1}, \qquad g^{0}_{2} = g^{}_2 \mu^{\varepsilon}Z_{g_2}, \qquad \tilde{\varphi} = \varphi Z_{\varphi} ,\nonumber \\
Z_1=Z_\varphi^2 ,  \qquad      Z_2=Z_{m^2} Z_\varphi^2, \qquad Z_3=Z_{g_1} Z_\varphi^4 ,\qquad Z_4=Z_{g_2} Z_\varphi^4,
\end{gather}
where $g_{1}^{0}$, $g_{2}^{0}$, $m_0^2$ are the bare quantities. As a result, the renormalized action has the following form
\begin{equation}
S^R = \int d^D{x}\Bigg\{\frac{1}{2} \Big[Z_1\partial_{\mu}\varphi_{i \alpha}^{\dagger}\partial_{\mu}\varphi_{\alpha i}^{}+ Z_2m^2\varphi_{i \alpha}^{\dagger}\varphi_{\alpha i}^{}\Big] + \frac{1}{4!}\Big[Z_3 g_{1}T^{(1)}_{\alpha i, \beta j, \gamma k, \delta l} +Z_4 g_{2}T^{(2)}_{\alpha i,\beta j,\gamma k,\delta l} \Big] \varphi_{i\alpha}^{\dagger}\varphi_{j\beta}^{\dagger} \varphi_{\gamma k }^{}\varphi_{\delta l}^{} \Bigg\}, \label{H_r}
\end{equation}
where $\mu$ is the renormalization mass, $g_1$ and $g_2$ are dimensionless coupling constant. As is known within the $\varphi^4$ based field theory, the renormalization constants are determined from the ultraviolet (UV) finiteness condition for the following two- and four-point one-particle irreducible Green functions:
\begin{eqnarray}
&&\Gamma^{(2)}_{\alpha i, \beta j} = \Gamma^{(2)}\delta_{\alpha i, \beta j}, \qquad \Gamma^{(4)}_{\alpha i,\beta j,\gamma k,\delta l} = \Gamma^{(4)}_1 T^{(1)}_{\alpha i,\beta j,\gamma k,\delta l}+\Gamma^{(4)}_2 T^{(2)}_{\alpha i, \beta j,\gamma k,\delta l},\\
&&\Gamma_1^{(4)} = \frac{2 \left(T^{(1)}_{\alpha i, \beta j, \gamma k, \delta l}(1+m\,n)-T^{(2)}_{\alpha i, \beta j, \gamma k, \delta l}(m+n)\right)}{m(m^2-1)n(n^2-1)} \Gamma^{(4)}_{\alpha i, \beta j, \gamma k, \delta l},\label{projectors1}\\
&&\Gamma_2^{(4)} = \frac{2 \left(T^{(2)}_{\alpha i, \beta j, \gamma k, \delta l}(1+m\,n)-T^{(1)}_{\alpha i, \beta j, \gamma k, \delta l}(m+n)\right)}{m(m^2-1)n(n^2-1)} \Gamma^{(4)}_{\alpha i, \beta j, \gamma k, \delta l}.
\label{projectors2}
\end{eqnarray}

The Minimal Subtraction (MS) scheme is used to renormalize the model. The counterterms in this scheme have the form:
\begin{equation}
Z_i(g_1,g_2,\varepsilon) = 1+\sum_{k =1}^{\infty} Z_i^{(k)}(g_1, g_2)\;\varepsilon^{-k}.
\label{Zi}
\end{equation}
The counterterms $Z_i$ depend only on $\varepsilon$ and coupling constants $g_1, \, g_2$. Bogoliubov-Parasyuk $R'$ operation could be used in order to calculate the renormalization constants. Thus renormalization constants can be expressed in terms of Green functions as follows.
\begin{equation}
Z_i=1 + KR' \bar{\Gamma}_i,
\end{equation} 
where $R'$ -- incomplete Bogoliubov-Parasyuk $R$-operation, $K$ -- projector of the singular part of the 
diagram and $\bar \Gamma_i$ -- normalized Green functions of the basic theory (see e.g. \cite{Vasilev,BogShirk}) defined by the following relations:
\begin{gather}
\bar{\Gamma}_1=\frac{\partial}{\partial {m^2}}\Gamma^{(2)}\mid_{p=0},  \quad \bar{\Gamma}_2=\frac{1}{2}\left(\frac{\partial}{\partial p}\right)^2\Gamma^{(2)}\mid_{p=0}  \quad \bar{\Gamma}_3=\frac{1}{g_1 \mu^{\varepsilon}}\Gamma^{(4)}_1\mid_{p=0}, \quad 
\bar{\Gamma}_4=\frac{1}{ g_2 \mu^{\varepsilon}}\Gamma^{(4)}_2\mid_{p=0}\;.
\end{gather}

The calculation of the counterterms for $\text{U}(n)\times\text{U}(m)$ symmetric model is based on computing the diagrams for O(1) symmetric (scalar) one. The counterterm values can be taken from the data obtained in the course of six-loop calculations for $\text{O}(n)$ symmetric model~\cite{KP17,BCK16,KompanietsPanzer:LL2016}. The information about tensor factors for particular diagrams can be calculated by apply projectors~\eqref{projectors1},~\eqref{projectors2} to it (see, e.g., Refs.~\cite{antonov2013critical,antonov2017critical,KalagovKompanietsNalimov:U(r)}). For this purpose, we use a tensor algebra package \textit{FORM}~\cite{Vermaseren:NewFORM} and manipulating graphs package \textit{Graphine/GraphState}~\cite{BatkovichKirienkoKompanietsNovikov:GraphState} which have proven to be sufficient for implementing the above-described approach when dealing with numerous diagrams.

\section{RG expansions and marginal dimensionalities}
\label{sec:rg_expansions}
In this section we define the $\beta$-functions and anomalous dimensions $\gamma_{\varphi}$ and $\gamma_m$. The way to extract the marginal dimensionalities is also presented here.

For the field theory with two coupling constants, the RG functions can be obtained employing the following expressions:
\begin{eqnarray}
\beta_i(g_1,g_2,\varepsilon) =\mu \frac{\partial g_i}{\partial\mu} \mid_{g_{01},g_{02}} = -g_i\left[\varepsilon -  g_1 \frac{\partial Z_{g_i}^{(1)}}{\partial {g_1}}-g_2 \frac{\partial Z_{g_i}^{(1)}}{\partial{g_2}}\right], \quad i = 1,2,\\
\gamma_j(g_1,g_2) = \mu \frac{\partial \log Z_j}{\partial \mu}\mid_{g_{01},g_{02}} = -  g_1 \frac{\partial Z_j^{(1)}}{\partial {g_1}}-g_2 \frac{\partial Z_j^{(1)}}{\partial{g_2}} , \quad j = \varphi, m^2,
\label{sl11}
\end{eqnarray}
where $Z^{(1)}_i$ are coefficients at the first pole in $\varepsilon$ taken from \eqref{Zi}. With the help of these relations, we calculate all the RG functions as expansions in terms of renormalized couplings up to six-loop contributions. Due to the lack of physical interest as well as the cumbersomeness of the expansions, we do not show them in the body of the paper. The information about full six-loop series for $\beta$ functions and anomalous dimensions $\gamma_{\varphi}$, $\gamma_{m^2}$ can be found in~\ref{app:sup_mat}. 

However, it is worth looking closer at some restrictions that can be extracted from the structure of the FP coordinates. In order to obtain them, it is necessary to solve the corresponding equation on zeros of $\beta$-functions:
\begin{equation}
\beta_1(g_1^*,g_2^*,\varepsilon)=0,\qquad \beta_2(g_1^*,g_2^*,\varepsilon)=0.
\label{beta0}
\end{equation}
By solving these equations iteratively, one finds the FP coordinates as series in $\varepsilon$. As stated above the $\text{U}(n)\times\text{U}(m)$ symmetric model may pose four FPs. We are interested only in $U^+$, which has the positive value of coordinate $g_2^*$, providing the proper symmetry breaking of QCD. 
The corresponding coordinates read like
\begin{eqnarray}
    g_{1,U^{\pm}}^{*}=\frac{36 - (m + n) (m + n \pm R)}{2 (108 + (m + n)^2 (-8 + m n))} \varepsilon, \qquad  g_{2,U^{\pm}}^{*}=\frac{5 n \mp 3 R + m (5 - n (m + n))}{108 + (m + n)^2 (-8 + m n)} \varepsilon,
\end{eqnarray}
where $R=\sqrt{24 + m^2 - 10 m n + n^2}$. The requirement $R\geq 0$ gives a restriction on the choice of a pair $(n, m)$. It is seen from the structure that within one-loop approximation, the values of parameters $n$ and $m$ can not be chosen independently. For one-loop approximation it can be written as $n^{\pm}=5 m \pm 2 \sqrt{6} \sqrt{m^{2}-1}+O(\varepsilon)$. For the smallest and physically nontrivial value of $m$ ($m=2$), this condition imposes the following restriction ($n\gtrsim 18.48$), which, of course, is an artifact of $\varepsilon$ expansion approach and lower orders of perturbation theory. In this regard, the higher orders are required to clarify the situation. It also demonstrates the motivation mentioned in the introduction to consider the generalized model instead of $U(n)\times U(n)$ symmetric one. The point is that in the case of parameters coincidence ($m=n$), only two FPs with negative $g_2$ exist when $n<\sqrt{3}+O(\varepsilon)$, unequivocally leading to the realization of the first-order phase transition for those systems which possess the positive value of bare coupling $g_2^0$. 

Thus, as was stated earlier, the numerical values of the marginal dimensionality $n^+$ for different $m$ are of the highest physical importance within the problem. The need to check the validity of the inequality $n^{+}(m,3)>m$ proves the conclusion about the realizability of a first-order phase transition for $U(n)\times U(n)$ symmetric models. At fixed $m$, the conditions of obtaining these quantities can be formulated as follows. In all known orders in $\varepsilon$, the following relations should be satisfied:
\begin{gather}
\beta_1(g_{1,\pm}^*(\varepsilon),g_{2,\pm}^*(\varepsilon),n^{\pm}(m,4-\varepsilon),\varepsilon)=0, \quad \beta_2(g_{1,\pm}^*(\varepsilon),g_{2,\pm}^*(\varepsilon),n^{\pm}(m,4-\varepsilon),\varepsilon)=0,\label{21} \\
\text{det}\left|\frac{\partial(\beta_1,\beta_1)}{\partial(g_1,g_2)} \right|(g_{1,\pm}^*(\varepsilon),g_{2,\pm}^*(\varepsilon),n^{\pm}(m,4-\varepsilon),\varepsilon)=0.
\label{22}
\end{gather}
From the equations above, the expansion for $n^+$ within six-loop approximation could be obtained in the following form: 
\begin{eqnarray}
 n^{\pm}(m)=n_{0}^{\pm}(m)+n_{1}^{\pm}(m) \varepsilon+n_{2}^{\pm}(m) \varepsilon^{2}+n_{3}^{\pm}(m) \varepsilon^{3}+n_{4}^{\pm}(m) \varepsilon^{4}+n_{5}^{\pm}(m) \varepsilon^{5}+\bigo{\varepsilon^6}.
\end{eqnarray}


\section{Numerical results}
\label{sec:numerical_results}
In this section, we calculate the $\varepsilon$ expansions of marginal dimensionality $n^+(m,4-\varepsilon)$ being the most physically important quantities for different values $m$. Following the equations~\eqref{21} and~\eqref{22} we obtain
\begin{eqnarray}
n^+(2,4-\varepsilon)&=&18.485-19.899\varepsilon + 2.926\varepsilon^2 + 4.619\varepsilon^3 - 0.718\varepsilon^4-1.766\varepsilon^5+\bigo{\varepsilon^6},\quad \label{eq:n_+_m_2}\\
n^+(3,4-\varepsilon)&=&28.856-30.083\varepsilon + 6.557\varepsilon^2 + 3.406\varepsilon^3 - 0.796\varepsilon^4-1.451\varepsilon^5+\bigo{\varepsilon^6},\quad \label{eq:n_+_m_3}\\
n^+(4,4-\varepsilon)&=&38.975-40.239\varepsilon + 9.609\varepsilon^2 + 3.050\varepsilon^3 - 0.616\varepsilon^4-1.370\varepsilon^5+\bigo{\varepsilon^6},\quad\label{eq:n_+_m_4}\\
n^+(5,4-\varepsilon)&=&49.000 - 50.375\varepsilon + 12.49\varepsilon^2 + 2.981\varepsilon^3 - 0.463\varepsilon^4 - 1.415\varepsilon^5+\bigo{\varepsilon^6},\quad\label{eq:n_+_m_5}\\
n^+(6,4-\varepsilon)&=&58.983 - 60.501\varepsilon + 15.29\varepsilon^2 + 3.043\varepsilon^3 - 0.342\varepsilon^4 - 1.515\varepsilon^5+\bigo{\varepsilon^6}.\quad\label{eq:n_+_m_6}
\end{eqnarray}
The series for $m=2,3,4$ are in full agreement with the known five-loop results~\cite{calabrese2004}. In order to obtain the proper numerical estimates in the case of $d=3$, the various resummation techniques combined with different algebraic manipulations will be applied to these expansions. 
\subsection{Resummation}
This section presents the numerical results of application the resummation techniques such as Pad\'e and Pad\'e-Borel-Leroy (PBL) to $\varepsilon$ expansions~\eqref{eq:n_+_m_2}-\eqref{eq:n_+_m_6}. A detailed explanation of these resummation methods can be found in \cite{kazakov1979analytic, kompaniets2016prediction, guida1998critical}, while the strategy for choosing the numerical values of the fitting parameters is described in~\cite{adzhemyan2019six}.
\begin{table}[b!]
\centering
\caption{Pad\'e estimates of marginal dimensionality $n^+(2,3)$. The numbers are obtained on the basis of initial  six-loop $\varepsilon$ expansion~\eqref{eq:n_+_m_2}.}
\label{tab:pade_nc_+_m2_6loop}
\renewcommand{\tabcolsep}{0.445cm}
\begin{tabular}{{c}|*{7}{c}}
$M \setminus L$ & 0 & 1 & 2 & 3 & 4&5 \\
\hline
0 & 18.4853 & -1.4142 & 1.5118 & 6.1303 & 5.4121 & 3.6460 \\
1 & 8.9021 & 1.1367 & -6.4724 & 5.5088 & 6.6225 & \text{} \\
2 & 6.0074 & 3.7049 & 4.6836 & 4.2501 & \text{} & \text{} \\
3 & 4.9506 & $\bold{4.2468}_6$ & $\bold{4.3749}_6$ & \text{} & \text{} & \text{} \\
4 & 4.5613 & $\bold{4.3371}_6$ & \text{} & \text{} & \text{} & \text{} \\
5 & 4.4234 & \text{} & \text{} & \text{} & \text{} & \text{} \\
\hline
\end{tabular}
\end{table} 
\begin{table}[t!]
\centering
\caption{
The PBL estimates of marginal dimensionality $n^+(2,3)$. The estimates are found by resummation of the initial $\varepsilon$ expansion~\eqref{eq:n_+_m_2}. The optimal value of resummation parameter $b_{opt}$ is $4.55$.}
\label{tab:pbl_nc_+_m2_6loop}
\renewcommand{\tabcolsep}{0.445cm}
\begin{tabular}{{c}|*{7}{c}}
$M \setminus L$ & 0 & 1 & 2 & 3 & 4&5 \\
\hline
0& 18.4853 & -1.4142 & 1.5118 & 6.1303 & 5.4121 & 3.6460 \\
1& 9.3145 & 1.0945 & -11.1470 & 5.5170 & 6.7700 & \text{} \\
2& 6.6398 & 3.4743 & $\bold{4.4190}_6$ & $\bold{4.4178}_6$ & \text{} & \text{} \\
3& 5.5904 & 4.0036 & $\bold{4.4178}_{6}$ & \text{} & \text{} & \text{} \\
4& 5.0979 & 4.1927 & \text{} & \text{} & \text{} & \text{} \\
5& 4.8394 & \text{} & \text{} & \text{} & \text{} & \text{} \\
\hline
\end{tabular}
\end{table}
Within the body of the paper, we only consider in detail the resummation steps for the case $m=2$ by discussing the algebraic manipulations and other ways to improve the convergence of numerical estimates for $n^{+}(2,3)$. The same steps were applied for $n^{+}(3,3)$ and $n^{+}(4,3)$. The corresponding results, including Pad\'e triangles and re-expanded expansions, are presented in~\ref{AppA}, while the final estimates are collected in Table~\ref{tab:n_+_final}. The numerical estimates will also be shown for $n^{+}(5,3)$ and $n^{+}(6,3)$ without any details.

First, one can address the method of Pad\'e approximants. The corresponding Pad\'e triangle for the series~\eqref{eq:n_+_m_2} is presented in Table~\ref{tab:pade_nc_+_m2_6loop}. The extracted result -- $4.32(7)$ -- is obtained based on the closest to each other approximants. The authors also demand the absence of poles for these approximants in $\varepsilon$ within the following range $[0,2]$. This choice of restricted radius from the physical value of $\varepsilon$ ($=1$) is conditional; it is dictated only by the computational experience of the authors and numerical considerations. 

In addition, the five loop counterpart was also obtained $4.2(5)$.

As is known, the Borel transformation (BT) can be applied to improve the convergence of the estimates. Despite the fact that the authors in~\cite{calabrese2004} claimed the uselessness of this procedure due to the irregular structure of the series~\eqref{eq:n_+_m_2}, by combining BT and the correct choice of resummation parameter values (Leroy parameter $b$), the authors of the present paper managed to obtain stable estimates. The corresponding Pad\'e triangle with PBL estimates for different approximants is presented in Table~\ref{tab:pbl_nc_+_m2_6loop}. We come to the following number $4.418(2)$ by means of the chosen strategy. The algorithmically obtained error bar, apparently, is underestimated. We consider at deriving the final result for $n^+(2,3)$, choosing the final value of inaccuracy as the standard deviation of the whole grid of numbers obtained using different resummation tactics. As previously, based on PBL technique, we also found the five-loop estimate $4.6(5)$ for $b = 35$. As is known, the parameter $b$ has no physical meaning, acting as a series convergence improver that allows it to be different for different orders.
\begin{table}[b!]
\centering
\caption{
Pad\'e estimates of marginal dimensionality $n^+(2,3)$.The estimates~\eqref{eq:n_+_m_2_a_full}  are found by resummation of the $\varepsilon$ expansion~\eqref{eq:n_+_m_2_a}.}
\label{pade_nc_+_m2_6loop_biased_part}
\renewcommand{\tabcolsep}{0.445cm}
\begin{tabular}{{c}|*{7}{c}}
$M \setminus L$ & 0 & 1 & 2 & 3 & 4&5 \\
\hline
0& 9.7426 & 4.1642 & 2.8380 & 4.4841 & 4.9481 & 4.2971 \\
1& 6.3372 & 2.4244 & 3.5725 & 5.1303 & 4.6772 & \text{} \\
2& 4.9795 & 4.2366 & 4.4555 & 4.3619 & \text{} & \text{} \\
3& 4.5511 & $\bold{4.3913}_6$ & $\bold{4.3847}_6$ & \text{} & \text{} & \text{} \\
4& 4.4384 &$\bold{ 4.3841}_6$ & \text{} & \text{} & \text{} & \text{} \\
5& 4.4021 & \text{} & \text{} & \text{} & \text{} & \text{} \\
\hline
\end{tabular}
\end{table}
\begin{table}[t!]
\centering
\caption{The PBL estimates of marginal dimensionality $n^+(2,3)$. The estimates~\eqref{eq:n_+_m_2_a_full} are found by resummation of the $\varepsilon$ expansion~\eqref{eq:n_+_m_2_a}. The optimal value of resummation parameter $b_{opt}$ is $10.2$.}
\label{pbl_nc_+_m2_6loop_biased_part}
\renewcommand{\tabcolsep}{0.445cm}
\begin{tabular}{{c}|*{7}{c}}
$M \setminus L$ & 0 & 1 & 2 & 3 & 4&5 \\
\hline
0& 9.7426 & 4.1642 & 2.8380 & 4.4841 & 4.9481 & 4.2971 \\
1& 6.4074 & 2.3610 & 3.5589 & 5.1577 & 4.6815 & \text{} \\
2& 5.1300 & 4.2365 & $\bold{4.3675}_{6}$ & $\bold{4.3675}_6$ & \text{} & \text{} \\
3& 4.6887 & 4.3426 & $\bold{4.3675}_6$ & \text{} & \text{} & \text{} \\
4& 4.5149 & 4.3613 & \text{} & \text{} & \text{} & \text{} \\
5& 4.4398 & \text{} & \text{} & \text{} & \text{} & \text{} \\
\hline
\end{tabular}
\end{table} 
In addition to the usage of various resummation techniques, in order to improve the convergence, some algebraic manipulations can be applied. Following the recipe suggested in~\cite{calabrese2004}, one can rewrite the expansions for $n^+(m,4-\varepsilon)$ as some biased series, taking advantage of the knowledge of $n^+(m,d)$ values in different spatial dimensionalities. In particular, the assumption that stable FP of two-dimensional model~\eqref{H} is equivalent to that of NL$\sigma$ model for all $n\geq1$ except $n=1$ turns out extremely fruitful~\cite{zinn1996book}. Combining with the fact of asymptotic freedom of NL$\sigma$ model, this allows authors in~\cite{calabrese2004} to conclude that $n^+(m,2)=1$~\cite{guida1998critical}. However, for the validity of this trick, one should assume the sufficient smoothness of $n^+(m,d)$ in $d$ at fixed $m$. Thus, the re-expanded series should be calculated in the following form:
\begin{equation}
 n^{+}(m, 4-\varepsilon)=1+(2-\varepsilon)\,a(m, \varepsilon), \label{eq:n_+_m_2_a_full}
\end{equation}
and in this case, the expansions for $a(m, \varepsilon)$ have to be resummed. On the basis of series~\eqref{eq:n_+_m_2} we obtain
\begin{equation}
a(2,\varepsilon) = 8.743 - 5.578\varepsilon - 1.326 \varepsilon^2 + 1.646 \varepsilon^3 + 0.464 \varepsilon^4-0.651\varepsilon^5+\bigo{\varepsilon^6}.\label{eq:n_+_m_2_a} \end{equation}
We follow the same steps as we did for the initial expansion~\eqref{eq:n_+_m_2}. The Pad\'e triangle for~\eqref{eq:n_+_m_2_a} in physical case ($\varepsilon=1$) is presented in Table~\ref{pade_nc_+_m2_6loop_biased_part}.
\begin{table}[b!]
\centering
\caption{Pad\'e estimates of marginal dimensionality $n^+(2,3)$. The numbers are obtained on the basis of the inverse six-loop $\varepsilon$ expansion~\eqref{eq:n_+_m_2_inv}.}
\label{pade_nc_+_m2_6loop_inverse_part}
\renewcommand{\tabcolsep}{0.38cm}
\begin{tabular}{{c}|*{7}{c}}
$M \setminus L$ & 0 & 1 & 2 & 3 & 4&5 \\
\hline
0&18.4843 & 8.90234 & 6.00745 & 4.9505 & 4.56142 & 4.42341 \\
1& -1.41421 & 1.13669 & 3.70494 &$\bold{ 4.24683}_6$ & $\textbf{4.33708}_6$ & \text{} \\
2& 1.51176 & -6.47249 & 4.68362 &$\bold{ 4.37484}_6$ & \text{} & \text{} \\
3& 6.13046 & 5.50873 & 4.25007 & \text{} & \text{} & \text{} \\
4& 5.41213 & 6.62252 & \text{} & \text{} & \text{} & \text{} \\
5& 3.64604 & \text{} & \text{} & \text{} & \text{} & \text{} \\
\hline
\end{tabular}
\end{table} 
\begin{table}[h!]
\centering
\caption{The PBL estimates of marginal dimensionality $n^+(2,3)$. The numbers are found on the basis of the inverse six-loop $\varepsilon$ expansion~\eqref{eq:n_+_m_2_inv}. The optimal value of resummation parameter $b_{opt}$ is $35$.}
\label{pbl_nc_+_m2_6loop_inverse_part}
\renewcommand{\tabcolsep}{0.38cm}
\begin{tabular}{{c}|*{7}{c}}
$M \setminus L$ & 0 & 1 & 2 & 3 & 4&5 \\
\hline
0& 18.4853 & 8.90234 & 6.00745 & 4.9505 & 4.56142 & 4.42341 \\
1& 1.30697 & 2.5455 & 3.25373 & 4.21443 &4.33088 & \text{} \\
2& 1.46921 & -16.415 & $\textbf{4.79065}_6$ & $\bold{4.38481}_6$ & \text{} & \text{} \\
3& 6.91898 & $\bold{5.88443}_6$ & 4.15679 & \text{} & \text{} & \text{} \\
4& 5.56979 & 10.1368 & \text{} & \text{} & \text{} & \text{} \\
5& -3.88727 & \text{} & \text{} & \text{} & \text{} & \text{} \\
\hline
\end{tabular}
\end{table} 
As seen, the numbers obtained by using different approximants are arranged closer to each other than those for the initial series~\eqref{eq:n_+_m_2}. Taking into account the unit shift, we come to $4.387(4)$ for Pad\'e estimate of $n^+(2,3)$. The five-loop counterpart was also obtained $4.36(9)$. The corresponding PBL triangle is presented in Table~\ref{pbl_nc_+_m2_6loop_biased_part}. As in the case of the Pad\'e triangle for $a(2,3)$, the numbers extracted based on different approximants turned out to be closer to each other than those for initial expansion~\eqref{eq:n_+_m_2}. Having analyzed the obtained results for series~\eqref{eq:n_+_m_2_a} we derive the following PBL estimate for $n^+(2,3)$: $4.36749(2)$. Of course, this error bar can not be perceived as genuine, reflected in the final estimate. The corresponding five-loop analysis gives $4.2488(9)$, but for different resummation parameter value ($b=2.25$). 

Also, to improve the convergence, the inverse series $1/n^+$ can be resummed \cite{calabrese2004, kompaniets2020six}. For $m=2$ we have
\begin{equation}
    \frac{1}{n^{+}(2,4-\varepsilon)}=0.0541 + 0.0582 \varepsilon + 0.0541\varepsilon^2 + 0.0355\varepsilon^3 + 0.0172\varepsilon^4 + 0.0068\varepsilon^5+\bigo{\varepsilon^6}.\label{eq:n_+_m_2_inv}
\end{equation}
The corresponding expansion coefficients decrease faster than those of the original series, which often should have a beneficial effect on the convergence. The six-loop results are $4.32(7)$ and $4.29(2)$ for Pad\'e and PBL techniques, respectively. Due to the rapidly diminishing coefficients of series~\eqref{eq:n_+_m_2_inv}, we also found the numerical result of its direct summation, which equals to $4.42$.

All the estimates for $n^{+}(2,3)$ obtained on the basis of the initial series~\eqref{eq:n_+_m_2}, the biased one~\eqref{eq:n_+_m_2_a}, and the inverse counterpart~\eqref{eq:n_+_m_2_inv} are collected in Table~\ref{tab:n_+_final}. As one can see, all the results are comparable with those obtained within previous order~\cite{calabrese2004}, increasing the accuracy of estimates. Note also that resummation of inverse expansions proved to be more or less successful only for $m=2$. At the same time, for other cases, we are not able, without resorting to some too counter-natural procedures, to extract the relatively stable numerical estimates. For this reason, we do not include the corresponding numbers in Table~\ref{tab:n_+_final}.

The sixth order of perturbation theory made it possible to improve the error by order of magnitude. All resummation details for different values of the parameter $m$ could be found in \ref{AppA}.

\begin{table}[H]
\centering
\caption{ Estimates of $n^+$ for several $m$ with varying the resummation techniques for $\text{U}(n)\times\text{U}(m)$ model 6 loops.}
\label{tab:n_+_final}
\renewcommand{\tabcolsep}{0.26cm}
\begin{tabular}{cc|ccccc}
\hline
        m          & &2  &3  & 4 & 5 &6\\ \hline
\multirow{2}{*}{initial series} & Pad\'e  & $4.32(7)$ & $7.213(7)$& $10.00(2)$ &$12.73(5)$ &$15.43(8)$ \\  
                  &PBL  & $4.418(2)$ & $7.19(4)$ &  $9.978(98)$& $12.7(2)$ &$15.4(3)$\\ \hline
\multirow{2}{*}{biased series} &Pad\'e   & $4.387(4)$ & $7.29(8)$ &  $10.08(5)$ &$12.81(3)$ &$15.51(1)$ \\  
                  & PBL &   $4.36749(2)$& $7.2269(2)$ &  $9.992(2)$& $12.708(1)$&$15.4269(2)$\\ \hline
  final estimate           &  & $4.373(18)$ & $7.230(22)$  & $10.012(28)$ & $12.74(05)$& $15.44(08)$\\ \hline
            \cite{calabrese2004}, 5 loop      &  &  $4.5(5)$ &$7.6(8)$  & $10.5(1.1)$ & - & -\\
            \hline
\end{tabular}
\end{table}

\section{Conclusion}

To summarize we calculated the six-loop RG expansions for the $\text{U}(n)\times\text{U}(m)$ symmetric model in $4-\varepsilon$ dimensions within the MS scheme. Together with different resummation techniques for these longer series allowed us to find yet the most accurate numerical results for marginal dimensionalities $n^+(m,3)$ for various $m$, that separate different regimes of critical behavior and determine the order of phase transitions in the $\text{U}(n)\times\text{U}(m)$ model. The accuracy of our estimates turns out to be one order of magnitude higher than those at five loops. These estimates are in favor of the inequality $m<n^{+}(m,3)$, supporting the feasibility of a first-order phase transition in light QCD.

\section{Acknowledgment}
M.V.K. and A.K. gratefully acknowledge the support of Foundation for the Advancement of Theoretical Physics "BASIS" through Grant 18-1-2-43-1.


\appendix

\section{Resummation \label{AppA}}
In this section, the results for $n^{+}(m,3)$ in case of $m=3, 4$ are presented. It was obtained by applying different resummation techniques, such as Pad\'e and Pad\'e-Borel-Leroy ones, to initial $\varepsilon$ expansions of the marginal dimensionalities $n^{+}(m, 4-\varepsilon)$ as well as to the corresponding biased series. The cases when $m>4$ are not considered in the paper, only the numerical estimates for $m=5,6$ are collected in Table~\ref{tab:n_+_final}. The rightness of the inequality $n^{+}(m,3)>m$ for $m>7$, however, has been verified by the authors.

\subsection{The numerical estimates of $n^{+}(3,3)$}

Let us again present the expansion for $n^{+}(3,4-\varepsilon)$:
\begin{equation}
n^+(3,4-\varepsilon)=28.856-30.083\varepsilon + 6.557\varepsilon^2 + 3.406\varepsilon^3 - 0.796\varepsilon^4-1.451\varepsilon^5+\bigo{\varepsilon^6}.\label{eq:n_+_m_3_app}
\end{equation} 
The Pad\'e triangle for series~\eqref{eq:n_+_m_3_app} is presented in Table~\ref{tab:appB_m3_pade_initial}. The obtained results are $7.66(43)$ and $7.213(6)$ for five- and six-loop approximations respectively. Using PBL technique we come to the number for $n^{+}(3, \varepsilon)$ at a sixth order as $7.19(4)$. The corresponding Pad\'e triangle with PBL estimates is present in Table~\ref{tab:appB_m3_pbl_initial}.
\begin{table}[h!]
\centering
\caption{Pad\'e estimates of marginal dimensionality $n^+(3,3)$. The numbers are obtained on the basis of initial  six-loop $\varepsilon$ expansion~\eqref{eq:n_+_m_3_app}.}
\label{tab:appB_m3_pade_initial}
\renewcommand{\tabcolsep}{0.395cm}
\begin{tabular}{{c}|*{7}{c}}
$M \setminus L$ & 0 & 1 & 2 & 3 & 4&5 \\
\hline
0 & 28.8564 & -1.2269 & 5.3297 & 8.7353 & 7.9395 & 6.4888 \\
1 & 14.1279 & 4.1564 & 12.4163 & 8.0902 & 9.7024 & \text{} \\
2 & 9.9431 & 6.6127 & 7.6795 & 6.6591 & \text{} & \text{} \\
3 & 8.3802 & $\bold{7.2207}_6$& $\bold{7.2097}_6$ & \text{} & \text{} & \text{} \\
4 & 7.7635 &$\bold{ 7.2094}_6$ & \text{} & \text{} & \text{} & \text{} \\
5 & 7.4822 & \text{} & \text{} & \text{} & \text{} & \text{} \\
\hline
\end{tabular}
\end{table} 
\begin{table}[h!]
\centering
\caption{The PBL estimates of marginal dimensionality $n^+(3,3)$. The estimates are found by resummation of the the initial $\varepsilon$ expansion~\eqref{eq:n_+_m_3_app}. The optimal value of resummation parameter $b_{opt}$ is $35$.}
\label{tab:appB_m3_pbl_initial}
\renewcommand{\tabcolsep}{0.395cm}
\begin{tabular}{{c}|*{7}{c}}
$M \setminus L$ & 0 & 1 & 2 & 3 & 4&5 \\
\hline
0 & 28.8564 & -1.2269 & 5.3297 & 8.7353 & 7.9395 & 6.4888 \\
1 & 14.2294 & 4.1352 & 12.9261 & 8.0928 & 9.8538 & \text{} \\
2 & 10.0887 & 6.5618 & 7.6566 & 6.7285 & \text{} & \text{} \\
3 & 8.5198 & $\bold{7.1757}_6$ & $\bold{7.2032}_6$ & \text{} & \text{} & \text{} \\
4 & 7.8678 & $\bold{7.2015}_6$ & \text{} & \text{} & \text{} & \text{} \\
5 & 7.5598 & \text{} & \text{} & \text{} & \text{} & \text{} \\
\hline
\end{tabular}
\end{table}

The corresponding biased expansion for $n^{+}(3, 4-\varepsilon)$ is
\begin{equation}
  a(3, \varepsilon) = 13.928 - 8.078 \varepsilon - 0.760 \varepsilon^2 + 1.323 \varepsilon^3 + 0.263\varepsilon^4 - 0.594\varepsilon^5+\bigo{\varepsilon^6}. \label{eq:n_+_m_3_a_app}
\end{equation} 
In Table~\ref{tab:appB_m3_pade_part} the Pad\'e triangle for $a(3, \varepsilon)$ is shown, by means of which we derive the five- and six-loop estimates for $n^{+}(3,3)$ as $7.43(28)$ and $7.29(8)$ respectively. Applying PBL method allows us to find the following numbers for $n^+(3,3)$: $7.2269(2)$ within six-loop approximation ($b = 0.65$), and $7.183(8)$ as its five-loop counterpart ($b = 0$). 
\begin{table}[h!]
\centering
\caption{Pad\'e estimates of marginal dimensionality $n^+(3,3)$. The numbers are obtained from the corresponding biased $\varepsilon$ expansion within six-loop approximation~\eqref{eq:n_+_m_3_a_app}.}
\label{tab:appB_m3_pade_part}
\renewcommand{\tabcolsep}{0.395cm}
\begin{tabular}{{c}|*{7}{c}}
$M \setminus L$ & 0 & 1 & 2 & 3 & 4&5 \\
\hline
0 & 14.9282 & 6.8507 & 6.0902 & 7.4128 & 7.6761 & 7.0825 \\
1 &  9.8157 & 6.0111 & 6.5730 & 7.7416 & 7.4937 & \text{} \\
2 &  8.0670 & 7.1857 & 7.4448 & $\bold{7.2006}_6$ & \text{} & \text{} \\
3 &  7.5259 & $\bold{7.3568}_6$ & $\bold{7.2979}_6$ & \text{} & \text{} & \text{} \\
4 &  7.3995 & 7.1657 & \text{} & \text{} & \text{} & \text{} \\
5 &  7.3185 & \text{} & \text{} & \text{} & \text{} & \text{} \\
\hline
\end{tabular}
\end{table} 
\begin{table}[h!]
\centering
\caption{The PBL estimates of marginal dimensionality $n^+(3,3)$. The numbers are found on the basis of six-loop $\varepsilon$ expansion~\eqref{eq:n_+_m_3_a_app}. The optimal value of resummation parameter $b_{opt}$ is $0.65$. \label{tab:appB_m3_pbl_part}}
\renewcommand{\tabcolsep}{0.395cm}
\begin{tabular}{{c}|*{7}{c}}
$M \setminus L$ & 0 & 1 & 2 & 3 & 4&5 \\
\hline
0 & 14.9282 & 6.8507 & 6.0902 & 7.4128 & 7.6761 & 7.0825 \\
1 & 10.4155 & 5.9709 & 6.5433 & 7.7679 & 7.5006 & \text{} \\
2 & 9.0424 & 7.1739 & 7.2297 & $\bold{7.2269}_6$ & \text{} & \text{} \\
3 & 8.4953 & 7.2247 & $\bold{7.2270}_6$ & \text{} & \text{} & \text{} \\
4 & 8.2247 & $\bold{7.2269}_6$ & \text{} & \text{} & \text{} & \text{} \\
5 & 8.0718 & \text{} & \text{} & \text{} & \text{} & \text{} \\
\hline
\end{tabular}
\end{table} 

\subsection{The numerical estimates of $n^{+}(4,3)$} 
The $\varepsilon$ expansion for $n^{+}(4, 4-\varepsilon)$ has the following form:
\begin{equation}
n^+(4,4-\varepsilon)=38.975-40.239\varepsilon + 9.609\varepsilon^2 + 3.050\varepsilon^3 - 0.616\varepsilon^4-1.370\varepsilon^5+\bigo{\varepsilon^6}.
\label{eq:n_+_m_4_app}
\end{equation}
The Pad\'e triangle for~\eqref{eq:n_+_m_4_app} in physical case ($\varepsilon=1$) is presented in Table~\ref{tab:appB_m4_pade_initial}. The resummed series allows us to extract for $n^{+}(4,3)$ the following numbers: $10.03(83)$ and $10.001(15)$ within the five- and six-loop  approximations respectively. The corresponding Pad\'e triangle with PBL estimates is shown in Table~\ref{tab:appB_m4_pbl_initial}. Using the chosen strategy, we obtain the following numbers: $10.5(12)$ and $9.978(98)$ for fifth and sixth orders respectively.
\begin{table}[h!]
\centering
\caption{Pad\'e estimates of marginal dimensionality $n^+(4,3)$. The numbers are obtained on the basis of initial  six-loop $\varepsilon$ expansion~\eqref{eq:n_+_m_4_app}.}
\label{tab:appB_m4_pade_initial}
\renewcommand{\tabcolsep}{0.395cm}
\begin{tabular}{{c}|*{7}{c}}
$M \setminus L$ & 0 & 1 & 2 & 3 & 4&5 \\
\hline
0 & 38.9737 & -1.2649 & 8.3436 & 11.3941 & 10.7786 & 9.4089 \\
1 & 19.1757 & 6.4915 & 12.8131 & 10.8819 & 11.8966 & \text{} \\
2 & 13.6660 & 9.2241 & 10.6020 & 8.6547 & \text{} & \text{} \\
3 & 11.5818 & $\bold{9.9835}_6$ & $\bold{10.0108}_6$ & \text{} & \text{} & \text{} \\
4 & 10.7427 &$\bold{ 10.0095}_6$ & \text{} & \text{} & \text{} & \text{} \\
5 & 10.3652 & \text{} & \text{} & \text{} & \text{} & \text{} \\
\hline
\end{tabular}
\end{table} 
\begin{table}[h!]
\centering
\caption{The PBL estimates of marginal dimensionality $n^+(4,3)$. The estimates are found by resummation of the the initial $\varepsilon$ expansion~\eqref{eq:n_+_m_4_app}. The optimal value of resummation parameter $b_{opt}$ is $35$.}
\label{tab:appB_m4_pbl_initial}
\renewcommand{\tabcolsep}{0.395cm}
\begin{tabular}{{c}|*{7}{c}}
$M \setminus L$ & 0 & 1 & 2 & 3 & 4&5 \\
\hline
0 & 38.9737 & -1.2649 & 8.3436 & 11.3941 & 10.7786 & 9.4089 \\
1 & 19.3122 & 6.4591 & \text{--} & 10.8837 & 11.9477 & \text{} \\
2 &\text{--} & 9.1654 & 10.5865 & 8.8001 & \text{} & \text{} \\
3 & 11.7666 & $\bold{9.9324}_6$ & $\bold{10.0045}_6$ & \text{} & \text{} & \text{} \\
4 & 10.8857 & $\bold{9.9958}_6$ & \text{} & \text{} & \text{} & \text{} \\
5 & 10.4714 & \text{} & \text{} & \text{} & \text{} & \text{} \\
\hline
\end{tabular}
\end{table} 

The biased part of $\varepsilon$ expansion for $n^{+}(4,4-\varepsilon )$ is 
\begin{equation}
  a(4, \varepsilon) = 18.987 - 10.626 \varepsilon - 0.509 \varepsilon^2 + 1.271 \varepsilon^3 + 0.328\varepsilon^4 - 0.521\varepsilon^5+\bigo{\varepsilon^6}. \label{eq:n_+_m_4_a_app}
\end{equation} 
Taking into account the unit shift, we come to $10.08(5)$ for Pad\'e estimate of $n^+(4,3)$. The five-loop counterpart is also obtained: $10.02(3)$. The corresponding Pad\'e triangle is presented in Table~\ref{tab:appB_m4_pade_part}. By means of PBL technique, the estimations of $n^{+}(4,3)$ are $9.96(18)$ and $9.992(2)$ within five- and six-loop approximations respectively. Table~\ref{tab:appB_m4_pbl_part} contains Pad\'e triangle with corresponding  PBL estimates.
\begin{table}[h!]
\centering  
\caption{Pad\'e estimates of marginal dimensionality $n^+(4,3)$. The numbers are obtained from the corresponding biased $\varepsilon$ expansion within six-loop approximation~\eqref{eq:n_+_m_4_a_app}.}
\label{tab:appB_m4_pade_part}
\renewcommand{\tabcolsep}{0.395cm}
\begin{tabular}{{c}|*{7}{c}}
$M \setminus L$ & 0 & 1 & 2 & 3 & 4&5 \\
\hline
0 & 19.9868 & 9.3610 & 8.8523 & 10.1232 & 10.4509 & 9.9299 \\
1 & 13.1738 & 8.8267 & 9.2156 & 10.5647 & 10.2497 & \text{} \\
2 & 10.9950 & 9.9021 & 10.2999 & 9.9590 & \text{} & \text{} \\
3 & 10.3166 & $\bold{10.1315}_6$ & $\bold{10.0789}_6$ & \text{} & \text{} & \text{} \\
4 & 10.1734 & $\bold{10.0234}_3$ & \text{} & \text{} & \text{} & \text{} \\
5 & 10.1007 & \text{} & \text{} & \text{} & \text{} & \text{} \\
\hline
\end{tabular}
\end{table} 

\begin{table}[H]
\centering
\caption{The PBL estimates of marginal dimensionality $n^+(4,3)$. The numbers are found on the basis of six-loop $\varepsilon$ expansion~\eqref{eq:n_+_m_4_a_app}. The optimal value of resummation parameter $b_{opt}$ is $0.1$.}
\label{tab:appB_m4_pbl_part}
\renewcommand{\tabcolsep}{0.395cm}
\begin{tabular}{{c}|*{7}{c}}
$M \setminus L$ & 0 & 1 & 2 & 3 & 4&5 \\
\hline
0 & 19.9868 & 9.3610 & 8.8523 & 10.1232 & 10.4509 & 9.9299 \\
1 & 14.2721 & 8.8125 & 9.1916 & 10.7591 & 10.2589 & \text{} \\
2 & 12.6370 & 9.8718 & 10.0214 & 9.9855 & \text{} & \text{} \\
3 & 11.9757 & $\bold{9.9908}_6$ & $\bold{9.9920}_6$ & \text{} & \text{} & \text{} \\
4 & 11.6458 & $\bold{9.9920}_6$ & \text{} & \text{} & \text{} & \text{} \\
5 & 11.4582 & \text{} & \text{} & \text{} & \text{} & \text{} \\
\hline
\end{tabular}
\end{table}
\section{Information about Supplemental Material}
\label{app:sup_mat}
In Supplemental Material we present expansions of RG functions for arbitrary values of $n$ and $m$. They ($\beta_1(g_1,g_2)$, $\beta_2(g_1,g_2)$, $\gamma_\phi(g_1,g_2)$, $\gamma_{m^2}(g_1,g_2)$) are put down as Mathematica file (\textit{rg\_expansions.m}). We also provide the Mathematica file with $\varepsilon$ expansions of marginal dimensionality $n^{+}(m,4-\varepsilon)$ under $m=\{2..6\}$ (\textit{marg\_dim\_exp.m}).
\bibliographystyle{elsarticle-num}
\bibliography{UnUm}
\end{document}